\documentclass[nofootinbib,superscriptaddress,preprint,longbibliography]{revtex4}
\usepackage{amsmath, amsthm, amssymb, amsfonts}
\usepackage{color}
\usepackage{graphicx}
\usepackage{epsfig}
\usepackage{longtable}

\begin{document}

\title{Constraints on the curvature of nuclear symmetry energy from recent astronomical data within the KIDS framework}
\author{Hana Gil}
\email{khn1219@gmail.com} 
\affiliation{Center of Extreme Nuclear Matter, Korea University, Seoul 02841, Korea}
\author{Panagiota Papakonstantinou}
\email{ppapakon@ibs.re.kr}
\affiliation{Rare Isotope Science Project, Institute for Basic Science, Daejeon 34047, Korea}
\author{Chang Ho Hyun}
\email{hch@daegu.ac.kr}
\affiliation{Department of Physics Education, Daegu University, Gyeongsan 38453, Korea}

\date{\today}

\begin{abstract}
We investigate the density dependence of the nuclear symmetry energy $S(\rho ) $ in the KIDS (Korea-IBS-Daegu-SKKU) framework for the nuclear equation of state (EoS) and energy-density functional (EDF). 
The aim is to constrain the value of the curvature parameter ($K_{\rm sym}$)  based on recent astronomical data. 
First, assuming a standard saturation point, we calculate bulk nuclear properties within KIDS-EDF for different values of the 
compression modulus of symmetric nuclear matter ($K_0$) and of the leading-order symmetry energy parameters, i.e., 
the symmetry energy ($J$) and slope ($L$) at saturation density, each within a broadly accepted range, as well as $K_{\rm sym}$. 
All of the above EoS parameters are varied independently of each other. 
The skewness parameter ($Q_{\rm sym}$) is presently kept fixed at 650 MeV. 
For all EoS parameter sets which describe the selected nuclear data within better than $0.3\%$, 
we calculate the neutron-star equation of state and mass-radius relation and analyze the results in terms of Pearson correlation coefficients $r$. 
We find that the value of $K_{\rm sym}$ is strongly correlated with the radius of both a canonical and a massive star ($|r|>0.9$). 
If we impose that all known astronomical constraints on the neutron star radii must be satisfied, we deduce $-150 < K_{\rm sym}<0$. 
As a result, the symmetry energy as a function of the density is consistently found to have an inflection point at $\rho_0<\rho<2\rho_0$. 
We take the opportunity to report that the neutron skin thickness of $^{208}$Pb shows no correlation at all with the neutron star radii ($|r|<0.1$), 
in contrast with studies which focus on the role of $L$ only.

\end{abstract}
\maketitle

\section{Introduction}

The physics of dense nuclear matter has entered a golden age thanks to previously unattainable data from diverse objects and phenomena in the Universe.
They include  millisecond pulsars~\cite{mass1,mass3}, X-ray bursts from low-mass X-ray binaries (LMXBs)~\cite{xburst,etivaux2019}, compact binaries~\cite{mass2}, gravitational waves (GWs) from neutron star merges~\cite{raithel2019,gw170817,gw2017high,gw190425}, and soft X-rays from pulsars~\cite{nicer1,nicer2021}. 
Most notable regarding the physics of neutron stars may be the possibility to determine the mass and radius of a single object, 
thus obtaining strong constraints to the mass and radius of neutron stars, and eventually the nuclear equation of state (EoS). 

One key issue for dense nuclear matter is to constrain the EoS and in particular the nuclear symmetry energy $S(\rho )$ at baryonic densities $\rho$ higher than the saturation density $\rho_0$ of symmetric nuclear matter~\cite{review2016,Oertel2017,lim2018,li2021}. 
By convention, the density dependence is encoded in the values of its derivatives at saturation density. 
The following expressions for the energy per particle $\mathcal{E}(\rho,\delta )$ at density $\rho$ and isospin asymmetry $\delta$ summarize the necessary definitions: 
\begin{eqnarray}
{\cal E}(\rho, \delta) &=& E(\rho) + S(\rho) \delta^2 + O(\delta^4), 
\label{Eq:EoSa}   \\[0.5mm]
E(\rho) &=& E_{0} + \frac{1}{2} K_0 x^2 + \frac{1}{6} Q_0 x^3 + \cdots, 
\label{Eq:EoSb}  \\[0.5mm]
S(\rho) &=& J + L x + \frac{1}{2} K_{\rm sym} x^2 + \frac{1}{6} Q_{\rm sym} x^3+ \cdots. 
\label{Eq:EoSc} 
\end{eqnarray}
Denoting the neutron and proton densities as $\rho_n$ and $\rho_p$, respectively, 
we have $\rho = \rho_n + \rho_p$, $\delta = (\rho_n -\rho_p)/\rho$, and $x = (\rho - \rho_0)/3 \rho_0$.
In symmetric nuclear matter ($\delta=0$) at densities where $|x|<1$, the density dependence of the EoS is
controlled dominantly by $K_0$, while in neutron-rich matter  ($\delta \approx 1$) the EoS below and above $\rho_0$ is largely driven by $L$ and $K_{\rm sym}$.

The saturation point as represented by the values of $E_0$ and $\rho_0$ is considered rather well known: Different theoretical assumptions lead to a maximal disagreement of the order of $1-2$~MeV for $E_0\approx -16$~MeV~\cite{atkinson2020} and $0.01 - 0.02$~fm for $\rho_0\approx 0.16$~fm$^{-3}$~\cite{dutra2012}, i.e., roughly 12$\%$ at most. 
The uncertainty in $J\approx 30$~MeV is of the order of a few MeV. 
Determining the EoS in the density region relevant for nuclei and neutron stars  is then equivalent to determining the values of $K_0$, $L$ and $K_{\rm sym}$. 
($Q_{\rm sym}$ is also a relevant parameter but with a more marginal role~\cite{kidsnm,kids_nuclei2}.) 
The above three parameters are poorly constrained.
$K_0$ is nonetheless best known among them: The empirically determined range is $200 - 260$ MeV~\cite{dutra2012,garg2018}.
A recent calculation based on microscopic theory predicts $K_0=282$~MeV~\cite{brandes2021}, which is one of the largest values in the recent literature.
The current status of $L$ is summarized comprehensively in Ref.~\cite{li2021}, 
where 24 new analyses 
of neutron stars give $57.7 \pm 19$ MeV at 68\% confidence level. 
The above range is consistent with the range $L = 58.7 \pm 28.1$ MeV provided by a previous survey \cite{Oertel2017} 
and with the earlier empirical range $L=40-76$~MeV~\cite{dutra2012}.

Compared with the above, $K_{\rm sym}$ is very poorly determined, with model estimates ranging from $-400$~MeV to positive values~\cite{li2021}.  
Let us summarize how current constraints on $K_{\rm sym}$ have been obtained. 
A standard way is to identify correlations between $K_{\rm sym}$ and better-known model parameters, namely $J$ and $L$, by surveying up to hundreds of models. 
A $K_{\rm sym}$ value is then deduced from the values of the other parameters.  
Recent works along this line propose $K_{\rm sym}$ values $-112 \pm 71$ \cite{Mondal2017}, and $-100 \pm 100$ MeV \cite{Mal2018}.
Another way to obtain correlations among EoS parameters is to use effective field theories that provide a platform to calculate
the correlations and uncertainties with realistic nuclear forces.
A recent work obtains $K_{\rm sym} = -150$ to $-70$ MeV \cite{lim2019}, which is a result of correlation between $K_{\rm sym}$ and $3J -L$ 
with appropriate ranges of $J$ and $L$.

One may also identify correlations between model predictions for observables that can be measured and the EoS parameter of interest.
One example is the relation of the energy of the giant monopole resonance with the nuclear incompressibility $K_0$~\cite{garg2018}. 
As regards $K_{\rm sym}$, recent works apply a Bayesian analysis to the tidal deformability of neutron stars.
The results vary broadly and depend on the technical details and assumptions. 
For example Ref.~\cite{Malik2018} obtains $K_{\rm sym} = -113$ to $-52$~MeV if $L=40-62$~MeV,
and Ref.~\cite{raithel2019} gives $K_{\rm sym} = -374$ to $45$~MeV.
The 16 new analyses of neutron star observables in \cite{li2021} suggest a range $K_{\rm sym} = -107 \pm 88$ MeV at 68\% confidence level.
This range excludes positive values, but the uncertainty is far larger than for $K_0$ and $L$. 

In a few cases $K_{\rm sym}$ is determined more directly from experiment.
In Ref. \cite{cozma2018}, elliptic flow data are used to obtain allowed ranges of $L$ and $K_{\rm sym}$.
The result is $K_{\rm sym} = 96 \pm 315 ({\rm exp}) \pm 170 ({\rm th}) \pm 166 ({\rm sys})$. 
While elliptic flow data are obtained from experiments on Earth, astronomical data have also been used: In Ref.~\cite{etivaux2019}, 
the thermal emission from LMXB is analyzed, and the authors obtain $K_{\rm sym} = -155$ to $-3$ MeV.  
The uncertaintly in $K_{\rm sym}$ is significantly lower than in the elliptic-flow work, but at the same time the value of $L$ was estimated to be $37.2 ^{+9.2}_{-8.9}$ MeV, which is on the lower side of the generally accepted range. 

Our poor knowledge of $K_{\rm sym}$ represents our poor knowledge of the EoS in the broad density domain necessary for the description of nuclei and neutron stars. 
Let us  point out also that most theoretical models providing input for the above analyses do not provide independent values for $K_{\rm sym}$, 
but rather values strongly correlated with the dominant parameters $J$ and $L$ owing to a deficiency in independent model parameters. For a discussion pertaining to Skyrme models see, e.g., \cite{prc103}.
As exceptions besides {\em ab initio} calculations we may mention the meta-model~\cite{Mal2018}, which however has minimal applicability at low densities and finite nuclei~\cite{meta_nuclei1}, and the KIDS 
(Korea-IBS-Daegu-SKKU) framework~\cite{kidsnm,kids_nuclei2,prc103} for the nuclear EoS and energy density functional (EDF), which provides a unified approach to both nuclear-structure and astronomical types of data and is used in this work.

The KIDS EoS of homogeneous matter, specifically the energy per particle,  has the analytical form 
\begin{equation} 
 \mathcal{E}(\rho,\delta ) = \mathcal{T}(\rho,\delta ) + \sum_{i=0}^{n}c_i(\delta ) \rho^{1+i/3} ,
\label{Eq:kidsnm} 
\end{equation} 
where $\mathcal{T}$ is the kinetic energy per particle of a free Fermi gas and $c_i(\delta ) $ can be assumed quadratic to $\delta$. The expansion in terms of the cubic root of the density is informed by effective field theories and {\em ab initio} calculations 
and can accommodate any set of EoS parameters as defined in Eqs.~(\ref{Eq:EoSb}),~(\ref{Eq:EoSc})~\cite{kidsnm,PP2018HNPS}. 
The EoS with the desired parameters is readily transposed to a nuclear EDF for finite nuclei with the additional freedom of choosing the values for the effective mass if desired~\cite{Gil2017,kids_nuclei1,kids_nuclei2}.
In Ref.~\cite{prc103}, we explored the density dependence of the symmetry energy in the KIDS framework by considering bulk properties of magic and open-shell nuclei 
and neutron-star properties. 
In particular, we explored the parameter space for $(J,L,K_{\rm sym})$ for a fixed EoS for symmetric matter, $E_0=-16$~MeV, $\rho_0=0.16$~fm$^{-3}$, $K_0=240$~MeV. The less relevant parameter $Q_{\rm sym}$ was fixed to $650$~MeV. 
We also examined $K_{\tau}$, which is the droplet-model counterpart of $K_{\rm sym}$ and more appropriate for nuclei, by using the correspondence 
\begin{equation}
K_\tau = K_{\rm sym} - \left( 6 + \frac{Q_0}{K_0} \right) L.
\label{eq:ktau}
\end{equation} 
Departures from existing model-dependent correlations between $K_{\rm sym}$ and $J,L$~\cite{Mondal2017} were observed. 
Among other results, we concluded that $K_{\rm sym}$ is negative and no lower than $-200$~MeV and suggested inconclusively $-400$ to $-300$~MeV as a range of likely values for $K_{\tau}$. 
In Ref.~\cite{npsm71}, the compression modulus $K_0$ of symmetric nuclear matter was explored in a similar manner and the results were found consistent with the current literature. 

In this work, we examine the acceptable values of $K_{\rm sym}$ by taking advantage of recent astronomical data from LMXB \cite{xburst}, GW170817 \cite{gw170817}, GW190425 \cite{gw190425} and NICER \cite{nicer2021}.
We explore both $K_0$ and $K_{\rm sym}$ at the same time, as well as $J$ and $L$. 
Note that in \cite{prc103}, $K_0$ is fixed to 240~MeV, while in Ref.~\cite{npsm71} the focus was on the uncertainties in $J$ and $L$. 
Considering broadly accepted ranges for the four parameters  $K_0,J,L,K_{\rm sym}$ (equivalently, $K_{\tau}$ instead of $K_{\rm sym}$) independently of each other, 
the total number of EoSs examined initially is about 4000. 
Next, we select a fraction of them (roughly$10\%$) which describe bulk data of selected nuclei most precisely, use them to calculate neutron-star properties, and analyze the results. 
We evaluate correlation coefficients for the model parameters and predictions, revealing a strong correlation between $K_{\rm sym}$ and neutron star radii. 
By imposing all astronomical constraints at the same time, we deduce $-150<K_{\rm sym}<0$~MeV.  
We take the opportunity to comment also on the neutron skin thickness of $^{208}$Pb measured recently by the PREX collaboration~\cite{prexII}. 

The paper is organized as follows. In Sec.~\ref{Sec:Nuclear} we perform the first filtering of parameter sets using bulk nuclear data. 
In Sec.~\ref{Sec:Astro}, we analyze our calculations for neutron-star properties and compare them with the above-mentioned astronomical constraints. 
The 358 sets used at this stage and selected results are available as supplementary material. 
We provide a summary and perspectives in Sec.~\ref{Sec:Sum}.

\section{EoS Parameters and correlation analysis\label{Sec:Nuclear}}
The standard KIDS EoS has three free parameters for the EoS of symmetric nuclear matter ($c_{i>2}(0)=0$) and four for neutron matter, or equivalently the symmetry energy ($c_{i>3}(1)=0$)~\cite{kidsnm,kids_nuclei2}. 
This means that seven leading EoS parameters, $(\rho_0,E_0,K_0)$ and $(J,L,K_{\rm sym},Q_{\rm sym})$, can be controlled independently. 
Higher-order parameters such as $Q_0$ do not vanish, but will fully depend on those seven.  
Although arbitrary extentsions to any number of free parameters are straightforward, 
the use of seven low-order KIDS parameters has been found optimal and suffices for our purposes. 
As already mentioned, compared to $K_0,J,L,K_{\rm sym}$, the saturation point $(E_0,\rho_0)$ is quite accurately known and will be kept fixed at $(-16$~MeV,~$0.16$~fm$^{-3})$. 
As in \cite{prc103}, the  less relevant parameter $Q_{\rm sym}$ will be kept fixed to the value of $650$~MeV. 

We begin by spanning the ranges $K_0 = [230, 260]$~MeV, $J=[30, 34]$~MeV, $L=[40, 70]$~MeV, $K_\tau = [-420, -240]$~MeV 
in increments of $\Delta J=0.5$~MeV, $\Delta K_0=\Delta L = 5$~MeV and $\Delta K_{\tau}=20$~MeV. 
In total, this gives us $7\times 9\times 7\times 10=4410$ EoS sets.  
At this stage, we use $K_{\tau}$ instead of $K_{\rm sym}$, because we are examining finite nuclei. 
The relation between the two is straightforward, Eq. (\ref{eq:ktau}). 
For the present choices, the value of $K_{\rm sym}$ ranges from $-300$ to $96$~MeV. 
The $Q_0$ value fully depends on $K_0$ and for the present choices ranges from $-373$ to $-313$~MeV.  

For each EoS, a KIDS-EDF is obtained, which has the form of an extended Skyrme functional~\cite{Gil2017,kids_nuclei1,kids_nuclei2,prc103}. 
Besides the EoS parameters, the necessary additional EDF parameters, namely $W_0$ (strength of spin-orbit force) and gradient terms represented by the Skyrme-like parameters $t_1, t_2$, are optimized by a fit to a few basic data~\cite{kids_nuclei2,Gil2017}. 
We note that the KIDS-EDF used at present has the simpler form given in Ref.~\cite{Gil2017}, i.e., the exchange momentum-dependent terms $y_{1,2}=t_{1,2}x_{1,2}$  are set to zero. 
The choice has no influence at all on our results for homogeneous matter, while any influence on bulk nuclear properties is too marginal to affect our current results~\cite{kids_nuclei1,prc103}.  

\begin{figure}
\begin{center}
\includegraphics[width=1.0\textwidth]{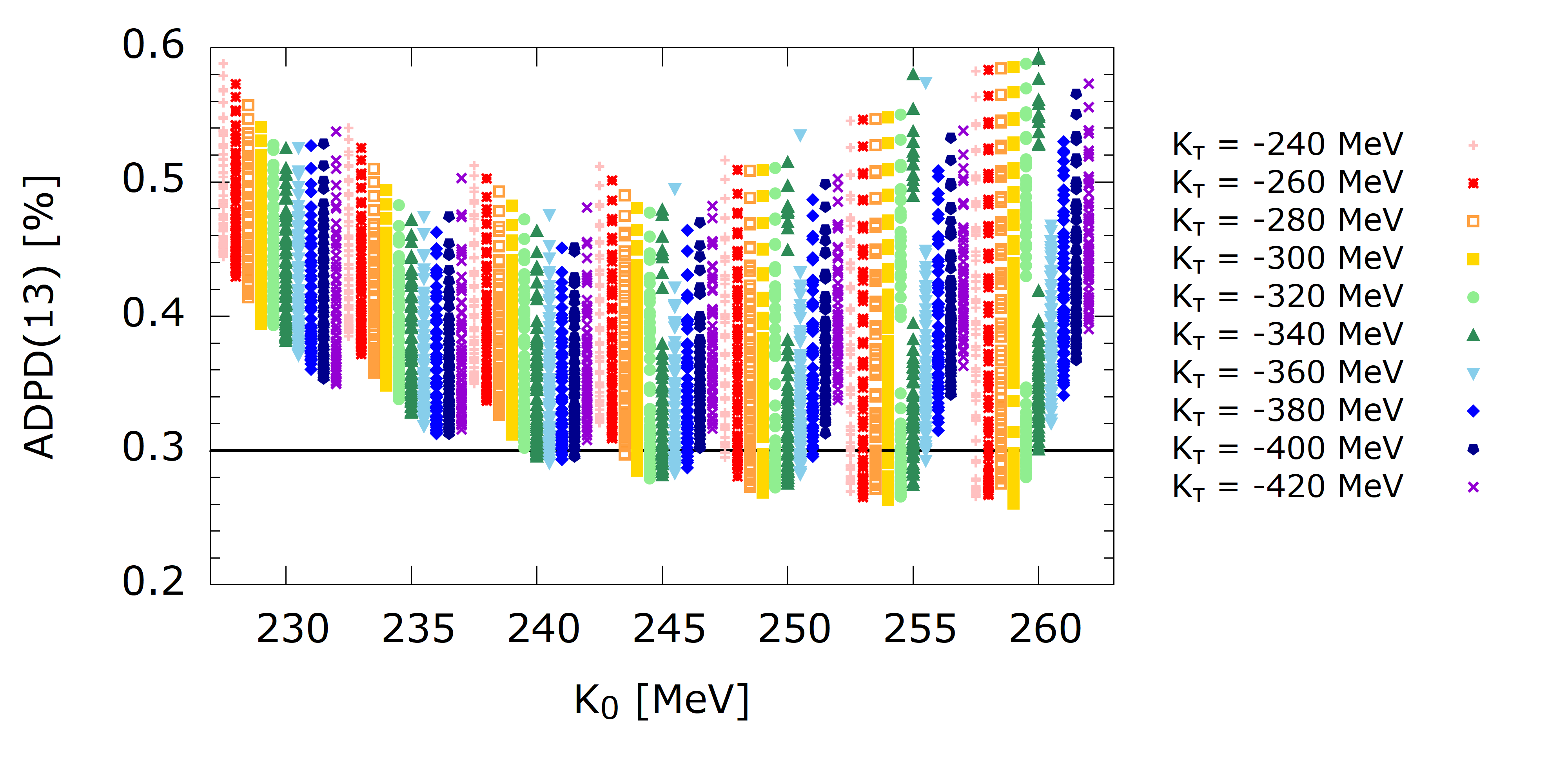}
\end{center}
\caption{Values of the criterion ADPD(13) with the examined sets of $(K_0, J, L, K_\tau)$. 
For each value of $K_0$, results with different $K_{\tau}$ are shown displaced for visibility. 
\label{Fig:adpd}}
\end{figure}

We then examine how accurately each EoS describes bulk nuclear properties by applying the corresponding EDF. 
As a measure of performance, we employ the average deviation per datum (ADPD) defined by
\begin{eqnarray}
{\rm ADPD}(N) \equiv \frac{1}{N} \sum^{N}_{i=1} \left| \frac{O^{\rm exp}_i - O^{\rm cal}_i}{O^{\rm exp}_i} \right|,
\label{Eq:adpd}
\end{eqnarray}
where $O^{\rm exp}_i$ and $O^{\rm cal}_i$ denote the value of an observable from experiment and theory, respectively.
Since the focus is on the EoS and not on nuclear structure, we will use only bulk properties of closed-shell nuclei. 
Specifically, we use $N=13$ data: binding energy~\cite{NNDC} and charge radius~\cite{Angeli2013} of 
$^{16}$O, $^{40}$Ca, $^{48}$Ca, $^{90}$Zr, $^{132}$Sn, $^{208}$Pb, and binding energy of $^{218}$U~\cite{NNDC}. 
The results are summarized in Fig.~\ref{Fig:adpd}. 
%

Roughly one tenth of all the EoSs give a value of ADPD less than $0.3\%$ and will be examined further.  
The values of their parameters are available as supplementary material. 
The cutoff of $0.3\%$ that we impose here  is a rather stringent one. 
Values for the same ADPD(13) as provided by representative Skyrme functionals are given in Table~\ref{tab:adpd}.  
\begin{table}
\begin{center}
\begin{tabular}{ccccc} \hline
 & SLy4 & HFB9 & SkM* & UNEDF0 \\[0.5mm] \hline
ADPD(13) & 0.86 & 0.87 & 0.84 & 0.98 \\[0.5mm]
ADPD(7)  & 0.30 & 0.36 & 0.38 & 0.59 \\[0.5mm]
\hline
\end{tabular}
\end{center}
\caption{ADPD(13), see Eq.~({\ref{Eq:adpd}}), in units of percentage, calculated with representative Skyrme force models.\label{tab:adpd}} 
\end{table}
Since those functionals are typically fitted to very different data, for comparison we also tabulate the values they give for 7 data only, namely the masses of $^{16}$O, $^{40}$Ca, $^{48}$Ca, $^{90}$Zr, $^{132}$Sn, $^{208}$Pb, and $^{218}$U.   

With each of the roughly 380 EoS sets selected, we calculate the neutron star EoS,
solve the TOV equations, and obtain the mass-radius relations of the neutron star in a standard way~\cite{kids_nuclei2,prc103}.
We are interested in particular in the radius $R_{1.4}$ and tidal deformability $\Lambda$ of a canonical neutron star with a mass roughly equal to $1.4$ solar masses and in the radius $R_{2.0}$ of a two-solar mass neutron star. All sets predict the existence of such massive neutron stars.

Next, we evaluate the Pearson correlation coefficient $r_{XY}$ between  EoS parameters or predicted observables $X,Y$
\begin{equation} 
r_{XY} = \frac{\overline{XY}-\bar{X}\bar{Y}}{\sigma_X\sigma_Y} ; \sigma_X = \sqrt{\overline{X^2} - {\bar{X}}^2} 
\label{Eq:corr} 
\end{equation} 
and list the results in the lower triangle of Table~\ref{tab:corr}. 
The values will be updated to those of the upper triangle after we consider the astronomical observations in Sec.~\ref{Sec:Astro}. 
We include the quantity $3J-L$, whose correlation with $K_{\rm sym}$ has been discussed before, see Ref.~\cite{prc103} and references therein. 
Along with EoS parameters, we include the predictions for $R_{1.4}$, $R_{2.0}$, and  the neutron skin thickness of $^{208}$Pb. 
\begin{table} 
\begin{tabular}{|c|rrrrrrrrr|} 
\hline 
                                   & $K_0$    &    $J$     &            $L$   &      \mbox{~}$3J-L$             & $K_{\rm{sym}}$   &  $K_{\tau}$  &    $\Delta R_{\rm np}$     &  $ R_{1.4}$  & $R_{2.0}$  \\[0.5mm] 
\hline 
   $K_0$                      &   1          &    $-0.51$    &     $-0.24$    &  \underline{$\mathit{0.10}$}   &   $0.35$         &      $0.50$     &         $-0.58$               &    $0.24$   &      $0.46$     \\[0.5mm] 
   $J$                          &  $-0.39$ &  1          &    $0.56$        &    $-0.35$       &         \underline{$\mathit{-0.06}$}             &     $-0.52$        &        $0.62$                 &     $0.22$      &    \underline{$\mathit{0.01}$}       \\[0.5mm] 
   $L$                           &  $0.24$  & 0.45     &         1          &      $\mathbf{-0.96}$   &  $0.50$            &       $-0.27$      &          $0.65$             &    $0.79$     &   $0.65$    \\[0.5mm]  
   $3J-L$                      &   $-0.34$    &   $-0.27$  &  $\mathbf{-0.98}$  &        1         &     $-0.60$       &   $0.14$    &   $-0.54$      &   $-0.84$    &    $-0.75$    \\[0.5mm] 
   $K_{\rm{sym}}$        &   $0.56$  &  \underline{$\mathit{0.09}$}   &      $\,\, 0.83$    &   $-0.88$  &           1                        &     $0.69$       &     $-0.29$         &     $\mathbf{0.90}$ &  $\mathbf{0.91}$    \\[0.5mm] 
   $K_{\tau}$               &  $0.62$   &  $-0.30$ &    $0.34$      &      $-0.44$       &   $ 0.80 $             &         1              &        $-0.85$            &      $0.33$      &   $0.43$      \\[0.5mm] 
   $\Delta R_{\rm np}$     & $-0.53$   &  $0.56$ &    $0.35$    &   $-0.26$    &    $-0.20$          &        $-0.70$        &       1                           &     \underline{$\mathit{0.09}$}       &       \underline{$\mathit{-0.05}$}     \\[0.5mm]   
   $ R_{1.4}$               & $0.52$   &  $0.21$ &    $\mathbf{0.91}$    &     $\mathbf{-0.93}$          & $\mathbf{0.96}$     &        $0.66 $        &         \underline{$\mathit{-0.02}$}                  &         1          &   $\mathbf{0.97}$    \\[0.5mm]   
   $R_{2.0}$                 & 0.63  &  \underline{$\mathit{0.09}$} &   $0.85$   &            $\mathbf{-0.90}$       &  $\mathbf{0.96}$          &        $0.69$        &     $-0.11$                     &    $\mathbf{0.99}$  &        1      \\[0.5mm]   
\hline 
\end{tabular} 
\caption{Correlation between the shown quantities according to Eq.~(\ref{Eq:corr}). Lower triangle: 
Considering sets satisfying ADPD(13)$<0.3\%$.  
Upper triangle: Considering sets which, in addition, satisfy selected astronomical constraints, see Sec.~\ref{Sec:Astro}. 
\label{tab:corr}} 
\end{table} 
Note that adding $Q_0$ to the table would offer no new information: Its correlation with all parameters would be the same as for $K_0$ and exactly 1 between $Q_0$ and $K_0$. 
That's because, with $\rho_0$ and $E_0$ fixed and having kept only three free parameters in the KIDS EoS of symmetric matter, $Q_0$ is fully and linearly determined by $K_0$. 
In addition, the correlation of $Q_{\rm sym}$ with the other parameters is not defined, because we have kept its value fixed.  

Numbers indicating strong (anti)correlation, $|r_{XY}|\geq 0.9$, are set in boldface, while those indicating absence of correlations, $|r_{XY}|\leq 0.1$, are set in underlined italics.   
Generally, the (anti)correlations between EoS parameters are weak at this stage, i.e., no linear correlations are apparent. 
Especially $K_{\rm sym} $ and $J$ appear uncorrelated. 
Neutron star radii are strongly correlated with $K_{\rm sym}$ and to some extent $L$ and $3J-L$. 
Notably, there are no correlations between the neutron skin thickness and the properties of neutron stars, 
in contrast with studies which focus on the role of $L$ only~\cite{pip2021}.

All the tabulated numbers should be interpreted with some care. The parameters and results which have been used as data sets do not necessarily follow gaussian or symmetric distributions, owing, for example, to the range cutoffs we have imposed 
in the EoS parameters, see also Fig.~\ref{Fig:adpd}. 
Also, different values would be obtained if different data and acceptance criteria were used. 
Nevertheless, the highest and lowest values of $|r_{XY}|$ marked in the table should carry some meaning. 
Also meaningful will be to inspect how the correlation patterns change or persist when we impose the neutron star constraints in the following.

\section{Results for neutron stars\label{Sec:Astro}}

For the canonical neutron star, $M= 1.4M_\odot$, the common range of all the data for $R_{1.4}$ from  LMXB, GW170817 and GW109425  is roughly $11.8 - 12.5$ km.
A recent work based on Bayesian analysis reports that the upper lmit of $R_{1.4}$ will not exceed $\sim 12.5$ km \cite{xburst2021}.
Therefore, we assume as acceptable a range of $R_{1.4} = 11.8 - 12.5$ km. 
For $R_{2.0}$ we only have the NICER measurement, taken for 2.08 solar masses, giving a much broader interval. 
\begin{figure}
\begin{center}
\includegraphics[width=1.0\textwidth]{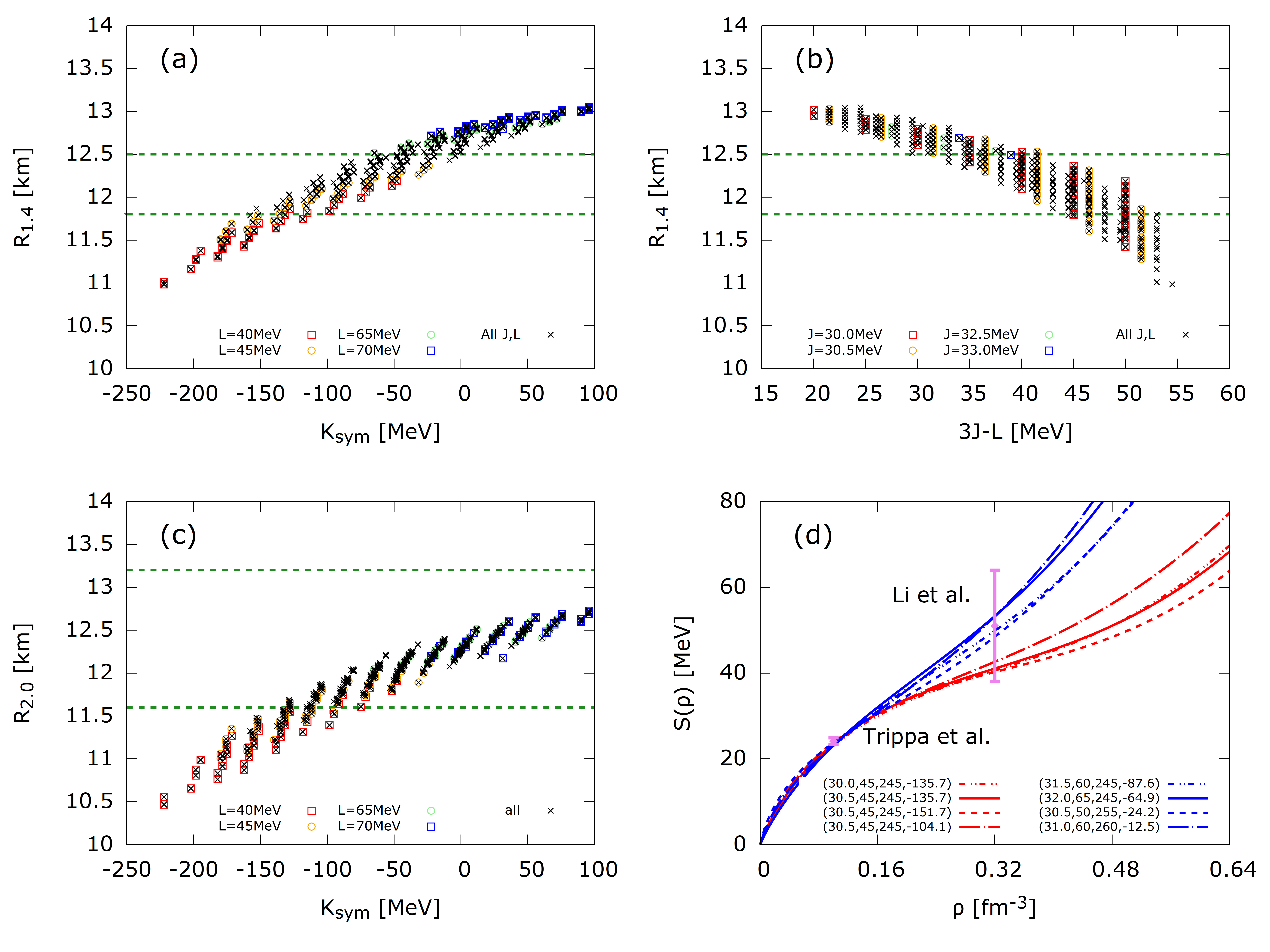}
\end{center}
\caption{(a)-(c): Predicted radii of neutron stars $R_{1.4}$ (for $1.4$ solar masses) and $R_{2.0}$ (for $2.0$ solar masses) 
shown against $K_{\rm sym}$ (a),(c) and $3J-L$ (b) for all sets satisfying the basic constraint from nuclear data, ADPD(13)$<0.3\%$.  
The dashed horizontal lines indicate combined astronomical constraints for the minimum and maximum acceptable values.\label{Fig:Astro}
(d): Symmetry energy corresponding to slected combinations of $(J,L,K_0,K_{\rm sym})$ values, indicated in units of MeV, roughly delimiting the allowed range based on the $R_{1.4}$ constraint in (a). 
The shown constraint at $0.1$~fm$^{-3}$ (Trippa {\em et al.}) is from Ref.~\cite{Trippa2008} and the estimate for $S(2\rho_0)$ (Li {\em et al.}) is from Ref.~\cite{li2021}.   }
\end{figure}

The results we obtained for neutron star radii  for the sets satisfying the basic constraint from nuclear data, ADPD(13)$<0.3\%$, are shown as scatter plots in Fig.~\ref{Fig:Astro} along with relevant astronomical constraints. 
Specifically, in Fig.~\ref{Fig:Astro} (a) and (c) we show plots of $R_{1.4}$ and $R_{2.0}$, respectively, against $K_{\rm sym}$ and in (b), a plot of $R_{1.4}$ against $3J-L$.  
Both $K_{\rm sym}$ and $3J-L$ were found strongly (anti)correlated with the radii (Table~\ref{tab:corr}, lower triangle). 
In order to reveal other possible systematics, in panels (a) and (c)  we highlight the results for the highest and lowest values of $L$
and in panel (b) we highlight values of $J$.  
All results 
are available as supplementary material. 

Imposing the strictest possible criterion of $11.8\leq R_{1.4}\leq 12.5$~km, as mentioned above and indicated in Fig.~\ref{Fig:Astro}(a), we obtain  
\[ 
   -150 \leq K_{\rm sym}  \leq 0 ~\mathrm{MeV} . 
\] 
Despite the varied range in $K_0$ and the different nuclear data considered in this work, our results for $K_{\rm sym}$ remain perfectly consistent with those in Ref.~\cite{prc103}, where we had deduced that $ -200 \leq K_{\rm sym}  \leq 0$ ~{MeV}. 
Let us note that we have also examined the dimensionless tidal deformability $\Lambda_{1.4}$ of a canonical neutron star, which we evaluated as $2.88\times 10^{-6} R_{1.4}^{7.5}$ (with $R_{1.4}$ in units of [km]) \cite{annala} and found strongly correlated with $K_{\rm sym}$.  
For the EoS sets which satisfy the imposed radius constraints, one finds $315<\Lambda_{1.4}<486$, consistent with current estimates from the LIGO/Virgo measurements. 

A stratification with respect to $L$ is apparent in Fig.~\ref{Fig:Astro}(a) with lower radius values favoring generally lower values of $L$. 
Higher values of $J$ are not favored. 
Overall, imposing again the strictest possible criterion of $11.8\leq R_{1.4}\leq 12.5$~km, as indicated in Figs.~\ref{Fig:Astro}(b), 
gives us  $30<3J-L<53$~MeV with $J=30-32$~MeV, implying roughly that 
\[ L < 3J -30\mbox{MeV} <66~\mbox{MeV} \quad , \quad L > 3J-53~\mbox{MeV} > 37~\mbox{MeV}  .\] 

One may impose more lenient criteria to the constraint on $R_{1.4}$, given the uncertainty in all measurements. 
Then broader ranges would be obtained for the EoS parameters. 
What is clear from our analysis is that 
an accurate and precise measurement of a neutron star's mass would be instrumental in constraining the value of $K_{\rm sym}$ - more generally, $S(\rho )$ at $\rho>\rho_0$.   

In Fig.~\ref{Fig:Astro}(d), we plot the symmetry energy corresponding to selected combinations of $(J,L,K_0,K_{\rm sym})$ values, indicated in units of MeV, roughly delimiting the allowed range based on the $R_{1.4}$ constraint in (a).
The upper curves correspond to $R_{1.4}$ closer to the  lower bound of $11.8$~km while the lower curves correspond to $R_{1.4}$ closer to the upper bound of $12.5$~km. 
A crossing point is observed at roughly $\rho=0.1$~fm, which is roughly the average density in a heavy nucleus, consistently with other studies~\cite{review2016}.  
As indicated on the Figure, the value of $S(0.1~\rm{fm}^{-3})$ is consistent with a constraint derived  from giant dipole resonances~\cite{Trippa2008}.
The symmetry energy at two times the saturation density is obtained consistent with the range proposed in Ref.~\cite{li2021}, which was extracted from a variety of heavy-ion collision data and astronomical data. 

An important observation is that in all cases the symmetry energy shows and inflection point at some density above saturation, i.e., the curvature changes sign from negative to positive. 
This behavior is mathematically possible because the higher-order parameters such as $Q_{\rm sym}$ do not vanish in this scheme. 
The inflection point is found here consistently within the interval between $\rho_0$ and $2\rho_0$.

The above constraints are based solely on bulk nuclear properties and existing astronomical data. 
We take the opportunity to comment on the neutron skin thickness of the nucleus $^{208}$Pb, which was extracted recently from parity violating electron scattering by the PREX collaboration~\cite{prexII} and found larger than expected from earlier estimates and known systematics~\cite{warda2011,piek2021}. 
It is generally thought that a moderate value of $L$, such as exctracted also in this work, does not favor a thick neutron skin. 
However, presently we find a rather weak correlation of the neutron skin thickness with $L$ compared with a much stronger anticorrelation with $K_{\tau}$. 
A thorough investigation of $K_{\tau}$ is therefore called for and is currently in progress. 

\begin{figure}
\includegraphics[width=0.45\textwidth]{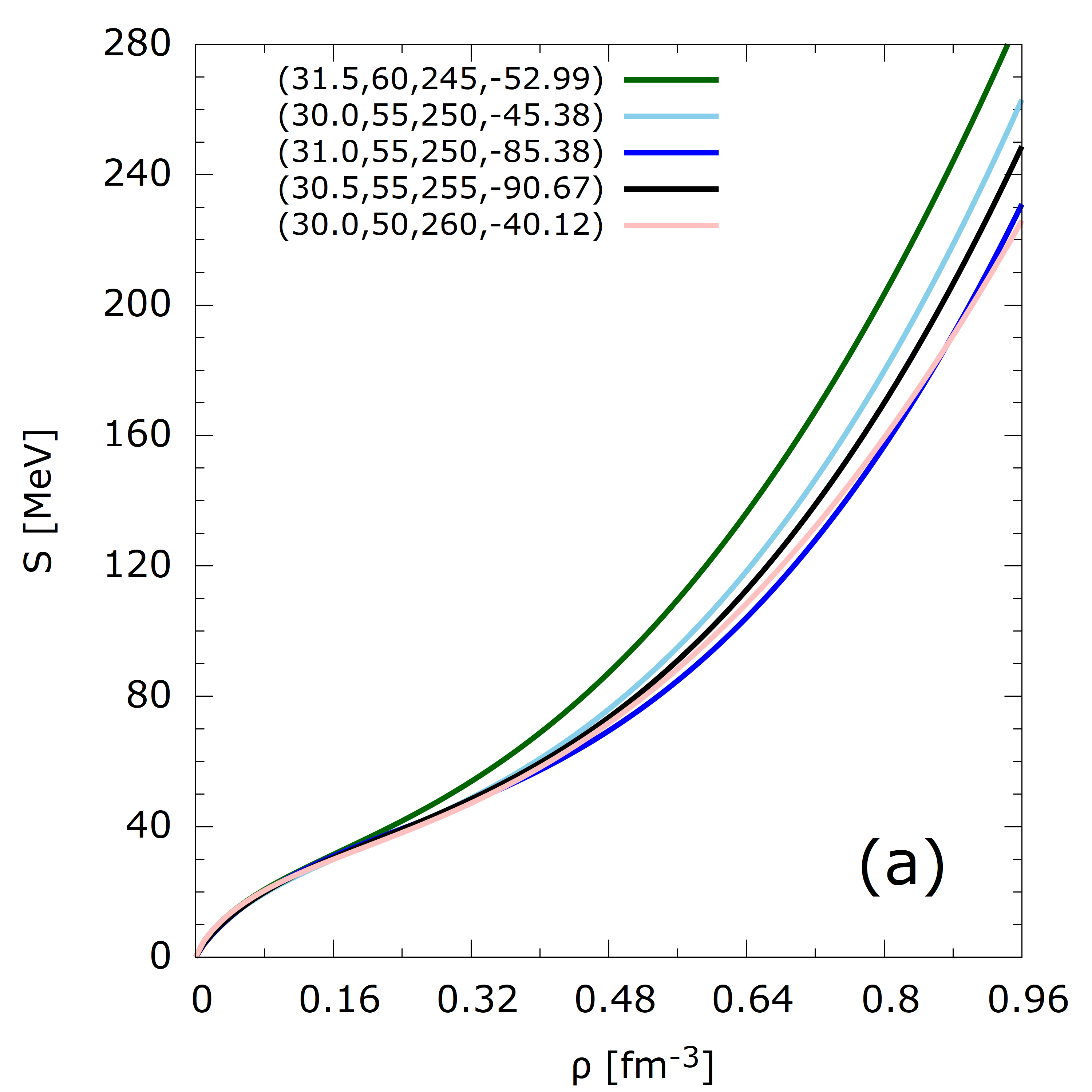}
\includegraphics[width=0.45\textwidth]{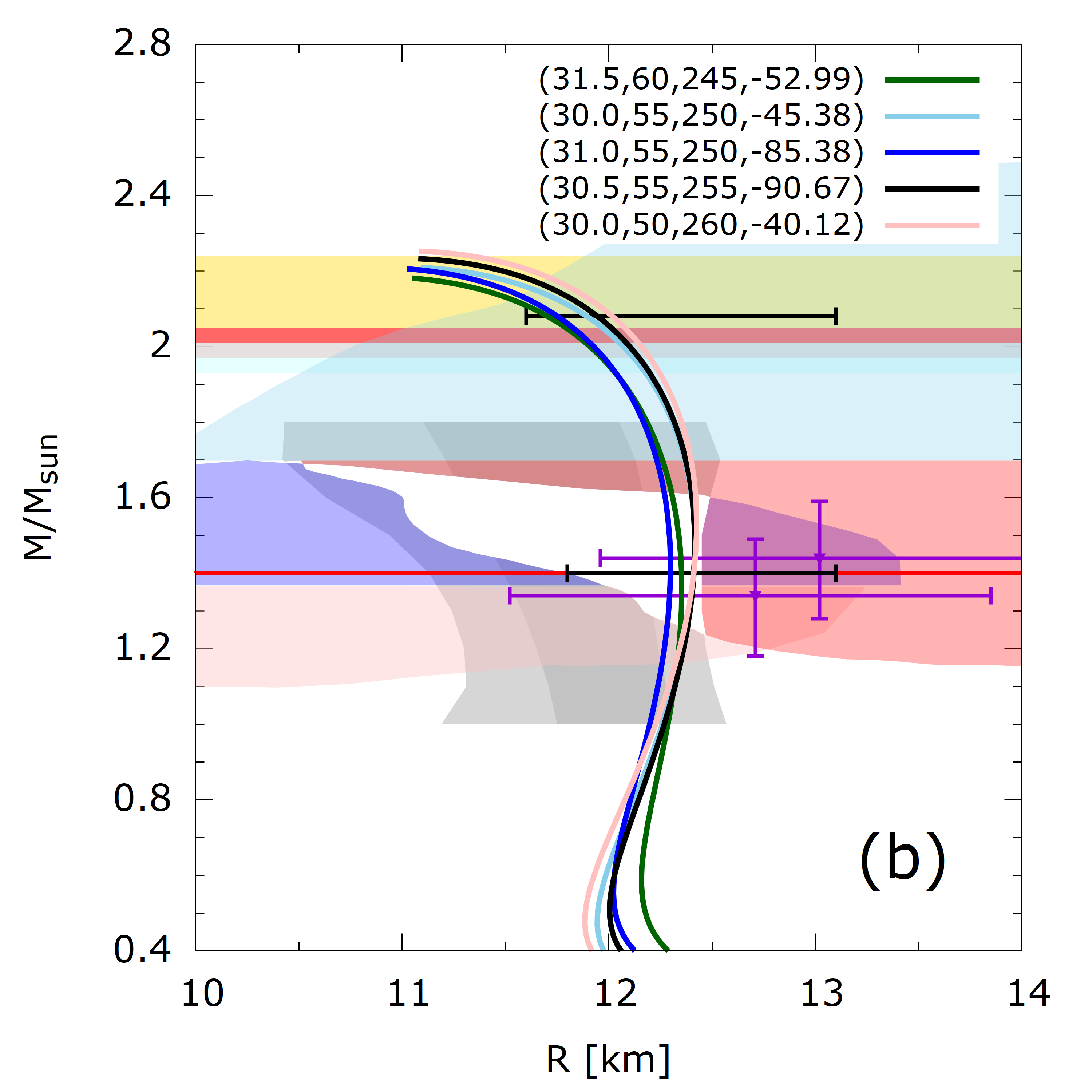}
\caption{Symmetry energy (a) and neutron star mass-radius curves (b) for representative EoSs satisfying astronomical and neutron-skin data as discussed in the text. 
The labels show the values of $(L,J,K_0,K_{\rm sym}$) in units of MeV. 
\label{Fig:EoSs}}
\end{figure}

Let us further inspect a few EoS sets obtained here which satisfy the astronomical constraints defined in Fig.~\ref{Fig:Astro} but also give a $\Delta R_{np}$ prediction for  $^{208}$Pb consistent with both the first PREX measurement, $0.33^{+0.16}_{-0.18}$~fm~\cite{prexI}, and an earlier broad estimate $0.195\pm 0.057$~fm~\cite{warda2011}. 
 In Fig.~\ref{Fig:EoSs} we plot the density dependence of the symmetry energy for five such sets (a) and the resulting neutron-star mass-radius relations (b). 
(One of the EoSs is consistent also with the PREX-II measurement by giving $\Delta R_{np}=0.212$~fm.)
Astronomical constraints from various sources are displayed in Fig.~\ref{Fig:EoSs}(b) related to masses observed or to masses and radii of neutron stars~\cite{mass3,mass2,xburst,gw170817,gw190425,nicer1,nicer2,nicerL28}. 
 What we conclude from this picture, compared to Fig.~\ref{Fig:Astro}(d), is that  combining both nuclear data (sub-saturation regime) and astronomical data (high-density regime) it becomes possible to constrain significantly the density dependence of the nuclear symmetry energy.
Precise astronomical data and more conclusive measurements of the neutron skin thickness are required.

\section{Summary\label{Sec:Sum}}

In this work, we attempt to constrain the curvature parameter of the symmetry energy $K_{\rm sym}$ based on bulk nuclear properties and available astronomical data. 
We utilize the KIDS framework for the nuclear EoS and EDF, which imposes no {\em a priori}, spurious correlations among EoS parameters. 
Assuming a standard saturation point, we calculate bulk nuclear properties within KIDS-EDF for different values of the compression modulus of symmetric nuclear matter ($K_0$) and of the 
leading-order symmetry energy parameters, i.e., the symmetry energy ($J$) and slope ($L$) at saturation density, each within a broadly accepted range, as well as  
$K_{\rm sym}$. 
The skewness parameter ($Q_{\rm sym}$) is presently kept fixed at 650~MeV. 
For all EoS parameter sets which describe the selected nuclear data within better than $0.3\%$, 
we calculate the neutron-star equation of state and mass-radius relation and analyze the results in terms of Pearson correlation coefficients $r$. 

The curvature parameter is found strongly correlated with neutron-star radii. 
By imposing that available measurements of neutron star radii must be reproduced within their given uncertainties, we arrive at an acceptable value of $K_{\rm sym}$ between roughly $-150$ and zero MeV. 
The symmetry energy consistently shows an inflection point at densities between $\rho_0$ and $2\rho_0$.  
Stronger constraints would be possible with more precise astronomical data. 
The neutron skin thickness of $^{208}$Pb shows no correlation at all with neutron star radii, but precise measurements would be useful for constraining the symmetry energy at sub-saturation densities, 
further aiding the evaluation of $K_{\rm sym}$.


\section*{Acknowledgments}
This work was supported by the National Research Foundation of Korea (NRF) grant funded by the Korea govenment
(No. 2018R1A5A1025563 and No. 2020R1F1A1052495)
and by the Rare Isotope Science Project of the Institute for Basic Science funded 
by the Ministry of Science, ICT and Future Planning and the National Research Foundation (NRF) of Korea (2013M7A1A1075764).


\mbox{~}
\newpage
\section*{Supplementary material} 

Values of parameters and results analysed  in this work are tabulated below. 
The EoS parameters $K_0,~J,~L,~K_{\rm sym}$ and the droplet parameter $K_{\tau}$ are given in units of MeV. 
The average deviation per datum ADPD(13) for 13 data (see Eq.~(6) of the main text) is given in percentage points \%. 
The spin-orbit and momentum-term couplings $W_0, t_1, t_2$ and the surface parameters 
$C_{12} = \frac{1}{64}[ -9t_1 + 5t_2 ]$, $D_{12}=\frac{1}{64}[3t_1+t_2]$ are given in units of MeV~fm$^{5}$. 
For the above parameters, typical values for the isoscalar and isovector 
nucleon effective masses are, respectively, 0.98 and 0.82 times the bare nucleon mass (with some deviations).
$M_1\approx 1.4M_{\odot}$ (more precisely in the range $(1.39-1.41)M_\odot$)
denotes the mass of a canonical neutron star and $M_2 = 2.0 M_\odot$ that of a massive star. 
The respective radii, $R_{M_1}$ and $R_{M_2}$ are given in units of km. 
The neutron skin thickness of $^{208}$Pb, $\Delta R_{np}$, is given in units of fm.

\footnotesize

\begin{longtable}{rrrrrrcrrrrrrrr} 
$K_0$  & $Q_0$  & $J$ & $L$ & $K_{\tau}$ & $K_{\rm sym}$  &  ADPD(13) & $W_0$  &  $t_1$ & $t_2$ & $C_{12}$ & $D_{12}$ & $R_{M_1}$ & $R_{M_2}$  & $\Delta R_{np}$ \\
\\[0.5mm] 240 	&	-372.65	&	30.5 	&	40	&	-340	&	-162.11 	&	0.298 	&	100.82 	&	293.18 	&	-172.48 	&	-54.70 	&	11.05 	&	11.44 	&	10.94 	&	0.164 
\\[0.5mm] 240 	&	-372.65	&	30.5 	&	40	&	-360	&	-182.11 	&	0.296 	&	98.44 	&	291.85 	&	-172.81 	&	-54.54 	&	10.98 	&	11.31 	&	10.83 	&	0.174 
\\[0.5mm] 240 	&	-372.65	&	31.0 	&	40	&	-340	&	-162.11 	&	0.297 	&	104.91 	&	294.50 	&	-172.28 	&	-54.87 	&	11.11 	&	11.43 	&	10.87 	&	0.163 
\\[0.5mm] 240 	&	-372.65	&	31.0 	&	45	&	-340	&	-139.87 	&	0.295 	&	102.06 	&	293.40 	&	-172.49 	&	-54.73 	&	11.06 	&	11.73 	&	11.22 	&	0.171 
\\[0.5mm] 240 	&	-372.65	&	31.0 	&	40	&	-360	&	-182.11 	&	0.291 	&	102.70 	&	292.57 	&	-172.84 	&	-54.65 	&	11.01 	&	11.30 	&	10.76 	&	0.173 
\\[0.5mm] 240 	&	-372.65	&	31.0 	&	45	&	-360	&	-159.87 	&	0.293 	&	99.58 	&	291.99 	&	-172.84 	&	-54.56 	&	10.99 	&	11.62 	&	11.14 	&	0.181 
\\
\end{longtable} 

\end{document}